\documentclass[	traditabstract]{aa}

\usepackage{natbib}
\usepackage{amssymb}
\usepackage{amsmath}
\usepackage{txfonts}
\usepackage{graphicx}
\usepackage{xspace}
\usepackage{subfigure}
\usepackage{wrapfig}
\usepackage{microtype}
\usepackage{paralist}
\usepackage{url}

\usepackage{lscape}

\newcommand{\chandra}{\textsl{Chandra}\xspace}
\newcommand{\fermi}{\textsl{Fermi}\xspace}

\newcommand{\inte}{\textsl{INTEGRAL}\xspace}
\newcommand{\sax}{\textsl{BeppoSAX}\xspace}
\newcommand{\suzaku}{\textsl{Suzaku}\xspace}
\newcommand{\xmm}{\textsl{XMM-Newton}\xspace}
\newcommand{\xte}{\textsl{RXTE}\xspace}

\newcommand{\msun}{\ensuremath{\text{M}_{\odot}}\xspace}
\newcommand{\rsun}{\ensuremath{\text{R}_{\odot}}\xspace}

\newcommand{\redchi}{\ensuremath{\chi^{2}_\text{red}}}
\newcommand{\cps}{\ensuremath{\mathrm{cts}\,\mathrm{sec}^{-1}}\xspace}
\newcommand{\nh}{\ensuremath{{N}_\mathrm{H}}\xspace}
\newcommand{\nhone}{\ensuremath{{N}_{\mathrm{H},1}}\xspace}
\newcommand{\nhtwo}{\ensuremath{{N}_{\mathrm{H},2}}\xspace}
\newcommand{\percmsq}{\ensuremath{\text{cm}^{-2}}\xspace}
\newcommand{\atmcmsq}{\ensuremath{\text{cm}^{-2}}\xspace}
\newcommand{\ergseccm}{\ensuremath{\mathrm{erg}\,\mathrm{sec}^{-1}\,\mathrm{cm}^{-2}}\xspace}
\newcommand{\gx}{GX~301$-$2\xspace}

\newcommand{\feka}{\ensuremath{\mathrm{Fe}~\mathrm{K}\alpha}\xspace}
\newcommand{\fekb}{\ensuremath{\mathrm{Fe}~\mathrm{K}\beta}\xspace}
\newcommand{\nika}{\ensuremath{\mathrm{Ni}~\mathrm{K}\alpha}\xspace}
\newcommand{\nikb}{\ensuremath{\mathrm{Ni}~\mathrm{K}\beta}\xspace}

\begin{document}

\title{Study of the many fluorescent lines and the absorption variability in \gx with \xmm }

\titlerunning{Fluorescent lines and absorption in  \gx}
\author{
 \mbox{F. F\"urst\inst{1}} \and
  \mbox{S. Suchy\inst{2}} \and
\mbox{I. Kreykenbohm\inst{1}} \and
\mbox{L. Barrag\'an\inst{1}} \and 
\mbox{J. Wilms\inst{1}} \and
 \mbox{K. Pottschmidt\inst{3,4}} \and
\mbox{I. Caballero\inst{5}} \and
\mbox{P. Kretschmar\inst{6}} \and
\mbox{C. Ferrigno\inst{7}} \and
\mbox{R. E. Rothschild\inst{2}}
}
\authorrunning{F.~F\"urst et al.}

\institute{
 Dr.~Karl Remeis-Sternwarte \& ECAP, Universit\"at Erlangen-N\"urnberg, Sternwartstr.~7, 96049~Bamberg, Germany
\and Center for Astrophysics \& Space Sciences, University of California, San Diego, 9500 Gilman Drive, La Jolla, CA 92093, USA
\and CRESST and NASA Goddard Space Flight Center, Astrophysics Science Division, Code 661, Greenbelt, MD 20771, USA
\and Center for Space Science and Technology, University of Maryland Baltimore County, 1000 Hilltop Circle, Baltimore, MD 21250, USA
\and CEA Saclay, DSM/IRFU/SAp --UMR AIM (7158) CNRS/CEA/Universit\'e P. Diderot, Orme des Merisiers Bat. 709, 91191 Gif-sur-
Yvette, France
\and ISOC, European Space Astronomy Centre (ESAC), PO Box 78, 28691 Villanueva de la Ca\~nada, Spain
\and ISDC data center for astrophysics, Universit\'e de Gen\`eve, chemin d'\'Ecogia, 16 1290 Versoix, Switzerland
}
\date{Received: --- / Accepted: ---}

\abstract{
We present an in-depth study of the High Mass X-ray Binary (HMXB) \gx during its pre-periastron flare using data from the \xmm satellite. The energy spectrum shows a power law continuum absorbed by a large equivalent hydrogen column on the order of $10^{24}$\,\atmcmsq and a prominent \feka fluorescent emission line. Besides the \feka line, evidence for \fekb, \nika, \nikb, S~K$\alpha$, Ar~K$\alpha$, Ca~K$\alpha$, and Cr~K$\alpha$ fluorescent lines is found. The observed line strengths are consistent with fluorescence in a cold absorber. This is the first time that Cr~K$\alpha$ is seen in emission in the X-ray spectrum of a HMXB. In addition to the modulation by the strong pulse period of $\sim$685\,sec the source is highly variable and shows different states of activity. We perform time-resolved as well as pulse-to-pulse resolved spectroscopy to investigate differences between these states of activity. We find that fluorescent line fluxes are strongly variable and generally follow the overall flux. The \nh value is variable by a factor of 2, but not correlated to continuum normalization. 
 We find an interval of low flux in the light curve in which the pulsations cease almost completely, without any indication of an increasing absorption column. We investigate this dip in detail and argue that it is most likely that during the dip the accretion ceased and the afterglow of the fluorescent iron accounted for the main portion of the X-ray flux. A similar dip was found earlier in \xte data, and we compare our findings to these results. 
}


\maketitle

\section{Introduction}
\label{sec:intro}

An X-ray flare at the coordinates of the  High Mass X-ray Binary (HMXB)  \gx was first detected in 1969 during a balloon experiment \citep{lewin71b}.  During further balloon flights \citet{mcclintock71a} found that the source was persistent. Since then \gx has been observed by most X-ray instruments and has stirred a lot of interest in the community as it shows a strongly variable flux and a very large absorption column. 
The source shows an erratically changing pulse period of $\sim\!685$\,sec, detected by \citet{white76a} and in the last few years monitored by  the \fermi/GBM  Pulsar Project \citep{finger09a}\footnote{see also \url{http://www.batse.msfc.nasa.gov/gbm/science/pulsars/lightcurves/gx301m2.html}}, clearly identifying the compact object as a neutron star.
The system has an orbital period of $P_\text{orb} =41.506\pm0.003$\,d at the reference time $T_0(\text{MJD}) = 43906.06 \pm 0.11$ and the orbital period is changing with $\dot{P}_\text{orb}= (-3.7\pm0.5)\times10^{-6}\,\mathrm{s}\,\mathrm{s}^{-1}$  \citep{doroshenko10a}. These values were determined by accurate pulse timing and are in agreement with earlier results from \citet{koh97a} who used the regular X-ray outburst of \gx. The eccentricity of the orbit was determined to be $e=0.47$.

The neutron star accretes matter from the optical companion, which was identified as the B-type giant Wray 977 \citep{vidal73a}. 
The exact type of the companion and its distance was long disputed, but the most recent measurements by \citet{kaper06a} indicate a distance of 3\,kpc and identify Wray~977 as a B1\,Ia+ hypergiant with a stellar-radius of 62\,\rsun and a mass of $43\pm10\,\msun$. We adopt these values throughout the paper, as X-ray accretion models by \citet{leahy08a} also seem to indicate similar values.

 The regular and bright X-ray flares of \gx were found to occur $\sim$1.4\,d before the periastron passage, i.e., before the neutron star reaches the densest part of the stellar wind. Standard, symmetric wind accretion models fail to explain pre-periastron flare. \citet{pravdo01a} suggested that a circumstellar disk is formed around Wray~977, from which the neutron star accretes additional matter when it passes through. If the disk is not aligned with the orbital plane of the neutron star, a pre-periastron flare could be observed. More recent calculations \citep{leahy02a,leahy08a} show that it is also possible that the neutron star is trailed by an accretion stream, which it overtakes shortly before periastron. This model self-consistently explains the observed dependence of flux and column density on the orbital phase. It is the regular pre-periastron flare which distinguishes \gx from most other HMXB.  During the flare the X-ray flux increases by up to a factor of 25 \citep{pravdo95a, rothschild87a}, reaching luminosities of up to 1\,Crab in the 15--50\,keV band and some 100\,mCrab in the softer 2--10\,keV band. X-rays with energies below $\sim$5\,keV are less variable over the orbit as they are strongly affected by scattering and absorption due to the high column density \citep{leahy08a}.

\begin{figure}
 \centering
 \includegraphics[width=0.95\columnwidth]{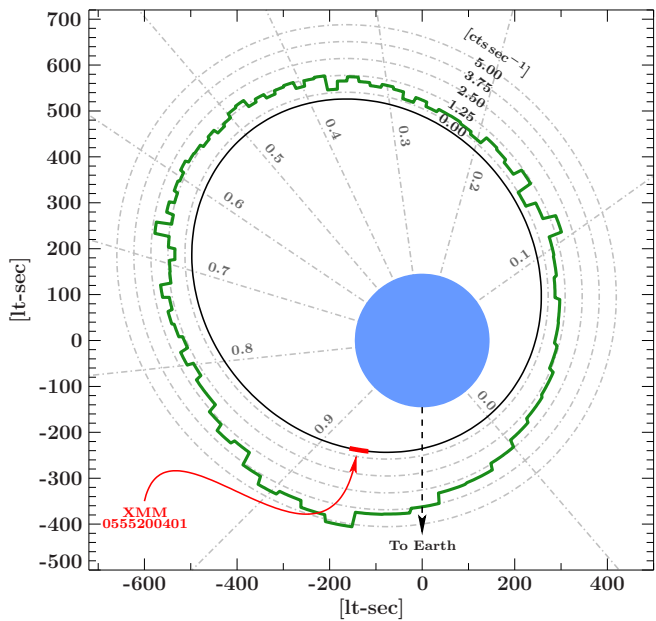}
 \caption{Sketch of the orbit of \gx, using the parameters given by \citet{koh97a}. The green line shows the average \xte-ASM light curve as function of orbital phase. The phase of the \xmm observation discussed here is highlighted in red.}
 \label{fig:orbit}
\end{figure}

Figure~\ref{fig:orbit} shows the orbit of \gx and the \xte-ASM light curve folded onto an orbital period of 41.5\,d using the ephemeris by \citet{doroshenko10a}. Clearly visible is the pre-periastron flare, peaking around phase $\varphi =0.91$. The plot assumes a size of $62\,\rsun$ for the optical companion \citep{kaper06a} . 

The broad-band X-ray spectrum of \gx can be described by a power law with a high-energy cutoff around 20\,keV and a cyclotron resonant scattering feature (CRSF) around 37\,keV \citep{endo02a}. \citet{kreykenbohm04b} showed that the spectral parameters, e.g., the energy of the CRSF, vary with pulse phase, and therefore concluded that the X-rays are most likely produced in an accretion mound close to the neutron star surface. In the softer X-rays a very strong \feka fluorescent emission line is visible together with a photo-electric absorption with an equivalent hydrogen column \nh of up to $10^{24}\,\atmcmsq$ \citep{leahy88a, leahy89a}. Using \chandra data, \citet{watanabe03a} detected a fully resolved Compton shoulder at the low-energy flank of the \feka line. These authors could model the shoulder by simulating the Compton scattering of photons in a dense, cold gas-stream.

The slow pulse period of \gx is strongly variable over long timescales \citep{pravdo01a, evangelista10a}.
 It was first detected with  $P\sim700$\,sec, but changed in the last 40\,years to values as low as $675$\,sec and is today around $P\sim 685$\,sec.
Although the period is variable, \gx usually shows clear pulsations throughout its orbit.
Recently, however, \citet{goegues11a}, discovered a dip in the \xte X-ray light curve during which \gx ceased to show pulsations. The data showed that exactly one pulse was missing and that the pulsed fraction was gradually reduced before  and gradually increased after the dip. Interestingly, the spectrum softened significantly during that dip, contradicting the common picture that a very dense blob, absorbing all soft X-rays, was passing through the line of sight. \citet{goegues11a} speculate that due to the cessation of accretion the black body radiation from the polar cap became visible, effectively softening the spectrum and reducing the pulsed fraction, as the polar cap is larger and undergoes stronger gravitational light bending than the accretion column. 
Similar dips have already been observed in the light curves of Vela~X-1 \citep{kreykenbohm08a}, but never before in \gx. 

The strong absorption during the bright pre-periastron flare makes \gx an ideal object to study the X-ray producing and absorbing regions around a neutron star, especially focusing on the many fluorescent lines. \xmm is the ideal instrument for this observation, with its large effective area and good spectral resolution. We obtained a $\sim$47\,ksec \xmm observation of which the results are presented here.

In the next Section we will describe the data used. Section\,3 will cover the timing analysis and in Sect.\,4 the different spectral analyses will be discussed. In Sect.\,5 the results are discussed and a conclusion and an outlook are given.

\section{Data \& Observations}
\label{sec:data}
We present data taken with the \xmm observatory on July 12th, 2009 (MJD 55024.103--55024.643, ObsID 0555200401) at orbital phase $\sim$0.91 of \gx, i.e., during the pre-periastron flare. The exposure was 46.6\,ksec, covering only a small part of the flare (see Fig.~\ref{fig:orbit}). The X-ray flux was strongly variable during the observation, with values between $3.67\times10^{-10}\,\ergseccm$ and $2.47\times10^{-9}\,\ergseccm$ in the 2--10\,keV energy band. 
Due to the expected high flux, the EPIC-MOS detectors were turned off to provide more telemetry for the EPIC-pn camera, which was operated in timing mode. In this mode the CCD is continuously read out, so that all spatial information in the read out direction is lost. In the other direction only the central 64 columns are used. The timing mode allows the measurement of countrates up to 400\,\cps without detectable pile-up
\citep{strueder01a}. As \gx shows very strong absorption no usable information can be gained from the Reflection Grating Spectrometers (RGS), which are mainly sensitive below 2.5\,keV.
We used the standard Science Analysis System (SAS) 10.0.0 to extract light curves and spectra. We tested for pile-up using the tool \texttt{epatplot}, but found no severe pile-up. The source data were extracted from EPIC-pn columns 35--40 and background data were obtained from columns 10--17 and 53--60.
As high photon fluxes change the charge transfer inefficiency (CTI) of the CCD detector, we also extracted spectra with the tool \texttt{epfast}, which takes these changes into account. We found, however, no significant differences to the standard pipeline.
The data were barycentric and binary corrected, using the ephemeris provided by \citet{doroshenko10a}.

\section{Timing Analysis}
\label{sec:timing}

\subsection{Light curve}

\begin{figure}
 \centering
 \includegraphics[width=0.95\columnwidth]{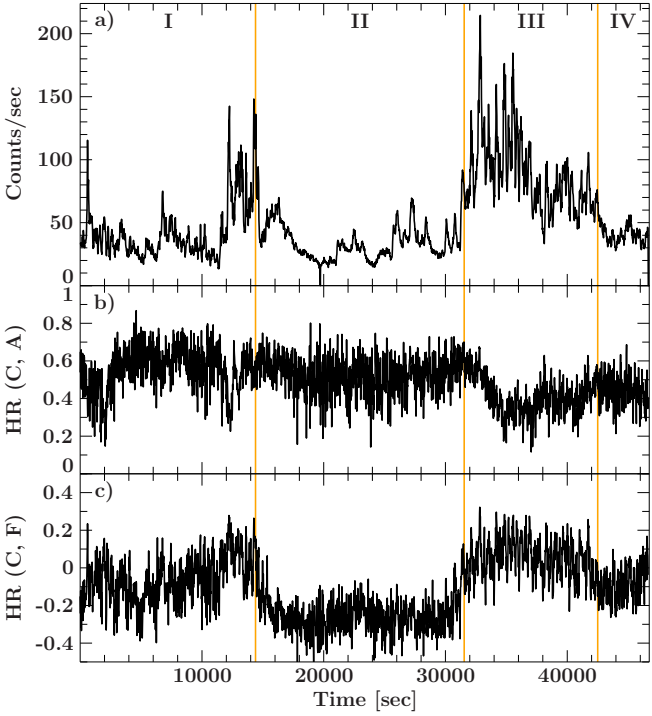}
 \caption{\textit{a)} Lightcurve of \gx with 20\,sec resolution in the 0.5--10.0\,keV energy band. Vertical lines indicate the time regimes I--IV used in the later analysis. \textit{b)} Hardness ratio between the continuum energy band $C$ (8.5--10.0\,keV) and the absorbed energy band $A$ (4.0--5.0\,keV). \textit{c)}   Hardness ratio between $C$ and the \feka line band $F$ (6.3--6.5\,keV).}
 \label{fig:lc}
\end{figure}

The top panel of Fig.~\ref{fig:lc} shows the 20\,sec resolution light curve of \gx. The light curve shows clear flaring behavior, with a smaller flare in the beginning and a large flare starting $\sim$30\,ksec after the start of the observation. Clearly visible are also the sub-peaks induced by the pulse period of $\sim$685\,sec. We performed a search for the pulse period via epoch folding and found a period of $P = 685.0 \pm 0.1$\,sec,
 consistent with the monitoring data from the \textsl{Fermi}/GBM Pulsar Project. 
In search for changes of the pulse period during the observation, we performed epoch folding and phase connecting on parts of the light curve as short as 15\,ksec. We did not find a significant change of the pulse period, consistent with \textsl{Fermi}/GBM data which shows that the pulse period was only slowly changing during that time.

\begin{figure}
 \centering
 \includegraphics[width=0.95\columnwidth]{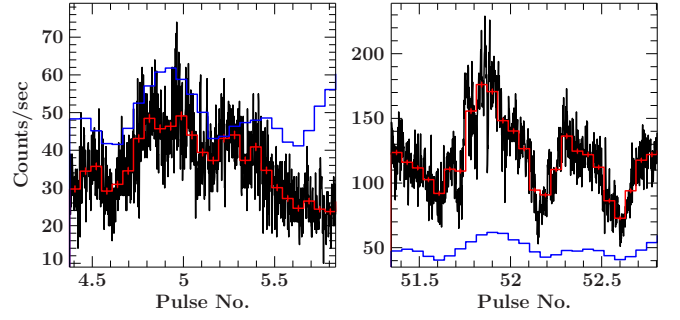}
 \caption{Two examples of the variability of the pulse shape, left an example early in the observation, right one close to the end of the observation. Shown in black is the 1\,sec light curve, in red a 20\,sec light curve. The blue profile is the average pulse profile over the complete observation. Note that the $y$-axes are not the same scale.}
 \label{fig:ppex}
\end{figure}

A close-up view of the light curve reveals that albeit the pulsations are clearly visible, strong pulse-to-pulse variations are also present. The individual pulse shapes range from simple two-peaked profiles to more complex shapes, with multiple minor peaks superimposed on the general shape. An example of two different profiles is shown in Fig.~\ref{fig:ppex} using light curves with 1\,sec and 20\,sec resolution. It becomes clear that the average pulse profile, also shown in Fig.~\ref{fig:ppex}, is highly complex. Additionally brightness changes independent of the pulses are visible in the light curve. These changes can be on the same order with respect to luminosity and duration as the variability due to the pulsation.

\begin{figure}
 \centering
 \includegraphics[width=0.95\columnwidth]{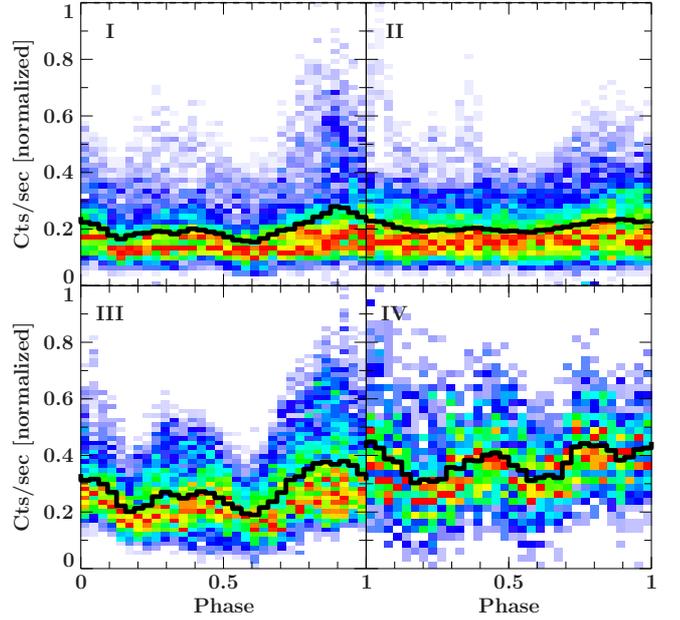}
 \caption{Color-coded maps of the pulse phase distribution of normalized countrates in regions I--IV as indicated in Fig.~\ref{fig:lc}. See text for details. Color represents the number of measurements for each countrate in the respective phase bin. Each phase bin was normalized to itself, so that they have all comparable color-codes. Colors are ranging from violet, corresponding to zero, to red, corresponding to one. The superimposed black line shows the mean value, i.e., the pulse profile.}
 \label{fig:pplsc}
\end{figure}

Stronger and sudden changes in luminosity are also seen in the light curve and different luminosity levels are marked as regions I--IV in Fig.~\ref{fig:lc}. These changes are also accompanied by changes in the pulse profile and the pulsed fraction. To investigate this in more detail, we folded each regime I--IV separately on the pulse period of $P = 685.0$\,sec and studied the normalized distribution of the countrate in each phase bin. As the overall countrate is different for every region we also normalized each light curve to its maximum countrate.
The mean of the distribution in each phase bin gives the value of the pulse profile in that bin. Figure~\ref{fig:pplsc} shows the results  for all four regimes in a color-coded map. Region III shows the clearest pulse profile, where the mean countrate as well as the whole distribution is changing with phase. This behavior is different in region I. Here the pulse profile is also visible in the mean values but the values measured most often are only weakly phase dependent. The two peaks of the profile result from strong outliers and flares at the respective phases. In region II a similar effect is seen, but the overall pulsations are even weaker. Region IV has a very short duration and therefore provides barely enough statistics for an interpretation.  The behavior seems similar to the one seen in region III, but the pulse profile is not that well pronounced.

\begin{figure}
 \centering
 \includegraphics[width=0.95\columnwidth]{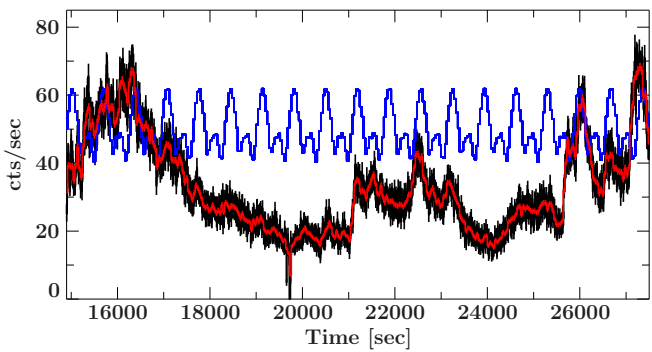}
 \caption{Small part of the light curve between 0.3--12.0\,keV, showing a vanishing of the pulsations. In black the light curve with 5\,sec resolution is shown, in red with 40\,sec resolution. Superimposed in blue is the average pulse profile for the whole observation.}
 \label{fig:vanishedpulse}
\end{figure}

Region II seems to be the region with the weakest pulsations and shows a very different behavior than region III.
To investigate these differences further, we took a closer look at the 1\,sec light curve during region II, as shown in Fig.~\ref{fig:vanishedpulse}. In the beginning the pulse is visible but it vanishes almost completely during the decline of the luminosity for $\sim$3 rotations. Around $t=21000$\,sec a sudden increase of the luminosity is accompanied by weak pulses but these disappear again after $\sim$4 rotations of the neutron star, also during a dip in the brightness.
Looking at the hardness ratio  during that time (Fig.~\ref{fig:lc}, lower two panels) allows us to deduce the spectral shape. The hardness ratio was calculated as $\text{HR}(H,S) =  (H -S)/( H +  S)$, with $H$ indicating the count rate in the harder and $S$ in the softer energy band.  When comparing the count rate in the 8.5--10\,keV continuum band $C$ to the count rate in the strongly absorbed energy band $A$ between 4--5\,keV, no significant variability of the hardness $\mathrm{HR}(C,A)$ can be measured during region II. Neither is a change compared to region I visible, indicating that the reduced flux is not due to increased absorption, which would primarily absorb the soft X-rays and would thereby make the spectrum overall harder. Contrary to that we see that the hardness $\text{HR}(C,F)$ between the count rate in the continuum band and in the iron line energy band $F$ between 6.3--6.5\,keV changes dramatically in region II. The strong drop of the hardness indicates that the flux of the iron line is not reduced as strongly as the continuum flux.

\subsection{Pulse profiles and pulsed fraction}
\label{susec:profiles}

\begin{figure}
 \centering
 \includegraphics[width=0.95\columnwidth]{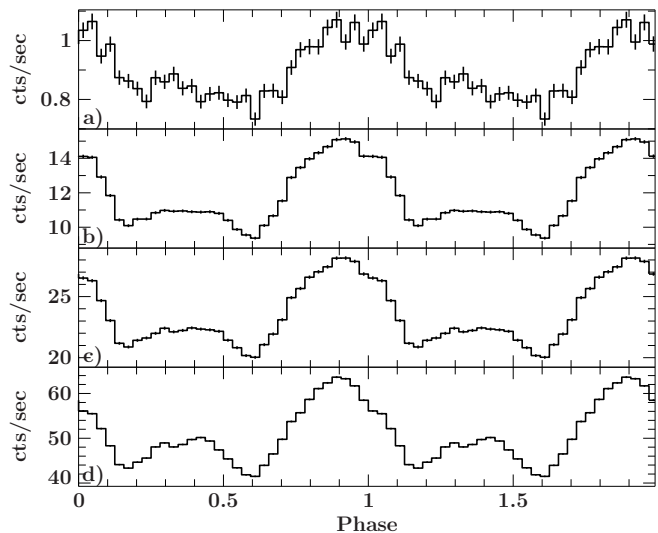}
 \caption{Pulse profiles in different energy bands: \textit{a)} 0.3--4.0\,keV, \textit{b)} 4.0--6.2\,keV, \textit{c)} 6.2--6.6\,keV, and \textit{d)} 6.6--15.0\,keV. Each profile is shown twice for clarity.}
 \label{fig:pulsprof}
\end{figure}

In the energy range of \xmm, the average pulse profile is only weakly energy dependent (Fig.~\ref{fig:pulsprof}). At higher energies the smaller, secondary peak becomes relatively stronger compared to the first one, but still stays clearly weaker.
Only at energies above 20\,keV the secondary peak gets almost as bright as the primary one \citep{kreykenbohm04b, labarbera05a}. This behavior is also confirmed in recent \suzaku data \citep{suchy11a}. 

\begin{figure}
 \centering
 \includegraphics[width=0.95\columnwidth]{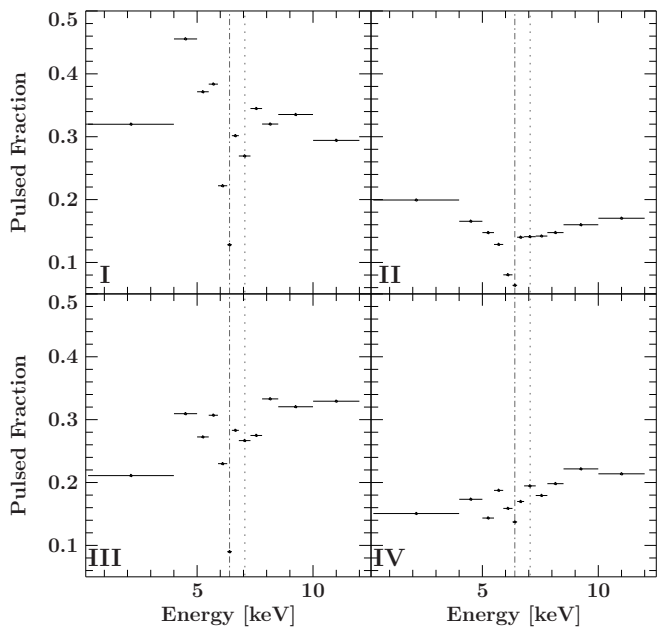}
 \caption{Pulsed fraction for different energy bands and for the 4 different regions I--IV. The dashed vertical line marks 6.4\,keV, the energy of the \feka line, the dotted one 7.1\,keV, the energy of the \fekb line.}
 \label{fig:pulfracenerg}
\end{figure}

\citet{endo02a} and \citet{tashiro91a} have shown that the pulsed fraction in the iron line energy band is significantly smaller than in other neighboring energy bands. Analysis of  \xmm data confirms this behavior, indicating that the matter producing the fluorescent iron line is distributed homogeneously around the X-ray producing region. The pulsed fraction~$f$ was calculated using the formula 
\begin{equation}
  f = \frac{\max(P)-\min(P)}{\max(P)+\min(P)}
\end{equation} 
with $P$ denoting the pulse profile.
The pulsed fraction is plotted in Fig.~\ref{fig:pulfracenerg} for the four different regions defined above. Common to all four regions is the fact that the pulsations in the iron line band are notably reduced. In region IV the dip is least prominent, probably due to the overall low pulsations and the short duration of this region.

Figure~\ref{fig:pulfracenerg} also clearly shows that the pulsed fraction is variable with time, dropping significantly during region II, as already seen in Fig.~\ref{fig:pplsc}. Some residual pulsed fraction is still measurable outside of the iron line band, but as the iron line band dominates the whole spectrum during that interval, almost no pulsations are visible in the broad band light curve. The lowest energy bin, between 0.5--4.0\,keV is dominated by the strong absorption, smearing out the pulsations and reducing the signal-to-noise ratio (S/N).
With increasing energy the pulsed fraction increases above 6.4\,keV, with exception of the energy bin covering the \fekb line at 7.1\,keV. The \fekb line is strong enough to contribute significantly to the flux in that bin, and the line is, as the \feka line, only very weakly pulsed. 

\section{Spectral analysis}
\label{sec:spectra}
\subsection{Phase averaged spectrum}
\label{susec:avg_spectra}

\begin{figure}
 \centering
 \includegraphics[width=0.95\columnwidth]{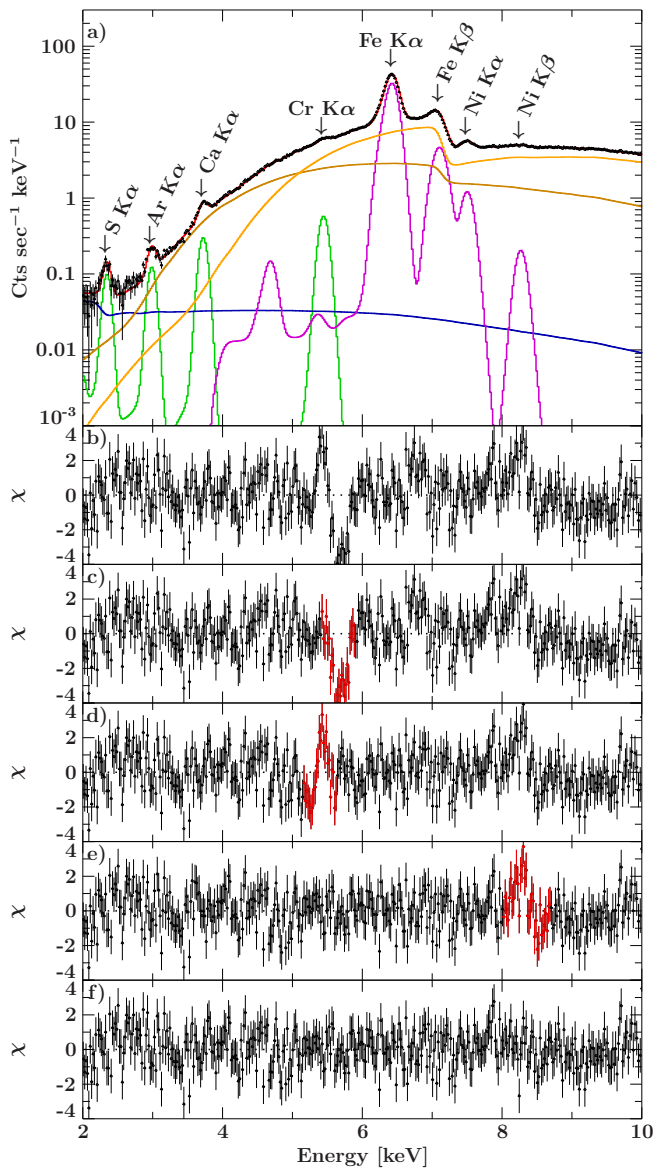}
 \caption{\textit{a)}: Phase and time averaged spectrum in the 2--10\,keV energy range of EPIC-pn. The components of the spectral model are shown in different colors, see text for a detailed description. \textit{b)} Residuals with neither an absorption feature nor an Cr K$\alpha$ line around 5.4\,keV. \textit{c)} Residuals with a Cr K$\alpha$ line, but without an absorption feature. \textit{d)} Residuals with an absorption feature, but without a Cr K$\alpha$ line. \textit{e)} Residuals without a Ni~K$\beta$ line. \textit{f)} Best fit residuals. }
 \label{fig:avgspec}
\end{figure}

When looking at the averaged X-ray spectrum of \gx (Fig.~\ref{fig:avgspec}\textit{a}) one sees a large iron K$\alpha$ fluorescent emission line and an equally large absorption column, which reduces the countrate to effectively zero below 2\,keV. The spectrum can be reasonably well modeled with a partially covered power law, with additional Gaussian lines at the \feka, Fe~K$\beta$, Ni~K$\alpha$, Ca~K$\alpha$, Ar~K$\alpha$, and S~K$\alpha$ fluorescent line energies.
We used a model of the form:
\begin{equation}
 \nh^\text{gal}  \times \biggl(\Bigl( c \nhone  + \left(1-c\right) \nhtwo \Bigr) \times \text{power law} + \text{Gaussian lines}\biggr)
\end{equation}
Where \nhone and \nhtwo are the two different parts of the partial coverer and $0 \leq c \leq 1$ is the covering fraction.
 We applied the \texttt{tbnew} absorption model (Wilms et al., 2011, in prep.)\footnote{see also \url{http://pulsar.sternwarte.uni-erlangen.de/wilms/research/tbabs/}} for both elements of the partial coverer.
For the large columns measured in \gx the plasma becomes optically thick and some flux is scattered out of the line of sight. This effect is taken into account with the \texttt{cabs} model, described by
\begin{equation}
 \text{\texttt{cabs}}(E) = \exp\Bigl(-\nh\,\sigma_\mathrm{T}(E)\Bigr)
\end{equation}
where $\sigma_\mathrm{T}(E)$ is the Thomson cross section and \nh is the equivalent hydrogen column. Both components of the partial coverer were multiplied by the \texttt{cabs} model using the same column densities, \nhone and \nhtwo, respectively.
The Gaussian components used to model the  fluorescent lines were only photo-electrically absorbed using a galactic $\nh^\text{gal} =1.7\times10^{22}\,\atmcmsq$ \citep{kalberla05a}, i.e., they were modeled as they appear outside the absorber.\footnote{DL value obtained  with  \texttt{nh} (\texttt{HEASOFT v6.10}), provided by HEASARC, at the coordinates of \gx} This description was chosen because the spectrum shows clear Ar and S~K$\alpha$ lines at 2.3 and 2.9\,keV, respectively, which can not be as strong as observed if they have to pass through an absorption column of $\sim\!10^{24}\,\atmcmsq$, as is necessary to describe the continuum. For a discussion on self-absorption and intrinsic strength of the lines, see Sect.~\ref{susec:flabund}.
This spectral model, including the same emission lines, can also describe the broad-band spectrum of \gx, as seen in recent \suzaku data \citep{suchy11a}.
Spectral parameters can be found in Table \ref{tab:4partpar} and a plot of the spectrum with the separate components and the residuals in Fig.~\ref{fig:avgspec}. In Table~\ref{tab:4partpar} \nhone gives the absorption column which is multiplied by the covering fraction (CF), while \nhtwo is multiplied by ($1-\mathrm{CF}$). The normalization and the photon index of the power law continuum are given in rows ``norm'' and ``photon index $\Gamma$'', respectively.

The \feka line shows strong residuals when fitted with a single Gaussian. \citet{watanabe03a} have shown that the \feka line shows an extended Compton shoulder (CS) at its lower energy side, but the resolution of EPIC-pn does not allow one to identify this feature uniquely.  \citet{bianchi02a} have shown that the Compton shoulder can be described with a box-shaped model. We included such a box in our model, which extends the Gaussian shaped \feka line to lower energies and starts at the centroid energy of the line. This model gives a very good description of the line, as evident in Fig.~\ref{fig:avgspec}. In Table~\ref{tab:4partpar} the row ``CS E$_1$'' gives the low energy end of the box, while ``CS A'' gives its area. The ratio of the flux between the \feka line and the Compton shoulder is in good agreement with the ratio expected for photons transmitted through a dense absorption column on the order of $10^{23}$\,\percmsq \citep{matt02a}.

The best fit $\chi^2$ obtained with this model was 769 with 376 degrees of freedom (dof) (\redchi = 2.04). Still snake-like residuals between 5.0--5.5\,keV were visible, see panel \textit{b)} of Fig.~\ref{fig:avgspec}. The K$\alpha$ line energy of neutral chromium is at 5.41\,keV \citep{kaastra93a, xraybooklet01}, and adding a Gaussian emission line in this energy range let to an improvement of the fit to $\chi^2 = 713$ with 374 d.o.f. The center of the line was found at  $5.425^{+0.005}_{-0.030}$\,keV in agreement with theory.
Still residuals remain around 5.6\,keV in absorption, see panel \textit{c)} of Fig.~\ref{fig:avgspec}. These residuals are  in the energy range of the plateau between the iron lines and their escape peaks in the silicon of the detector. The escape peaks and the plateau are clearly visible in the model of the Gaussian components of the \feka and \fekb line in Fig.~\ref{fig:avgspec} due to the large flux of these lines.
 The residuals could be very well described with a broad Gaussian absorption feature, improving the fit further to $\chi^2 = 488$ with 371 d.o.f. and called ``Abs. Feat.'' in Table~\ref{tab:4partpar}. Using only this absorption feature, and leaving out the Cr~K$\alpha$ line, to model the residuals does not adequately describe the spectrum, and gives a $\chi^2$ of only $562$ for 373 d.o.f., see Fig.~\ref{fig:avgspec}\textit{d}.

Finally strong residuals remained around 8.3\,keV, very close to the energy of the Nickel K$\beta$ line at 8.26\,keV  \citep{palmeri08a}. 
A Gaussian shaped line centered at that energy was used to model the residuals and improved the fit clearly.
This last addition resulted in our best fit model, with $\chi^2 =  448$ for 369 d.o.f. ($\redchi = 1.214$), of which the residuals are shown in Fig.~\ref{fig:avgspec}\textit{f}. 

To our knowledge, this is the first time that clear evidence for a Cr~K$\alpha$ as well as a Ni~K$\beta$ emission line was found in the soft X-ray spectrum of a HMXB. Cr has been observed in absorption in other sources before, e.g., in GRO~J1655$-$40 \citep{miller08a}. These authors found, however, no evidence for fluorescence lines in the spectrum of GRO~J1655$-$40 and Cr as well as all other Fe-like elements were found to be highly ionized.

Even though the spectrum was only used above 2\,keV, as all bins below are clearly background dominated, the continuum  in the energy range between 2--3\,keV can not be accurately described by only the partial covering model. A constant DC level is needed additionally, to lift the continuum in the lowest energy bins. Without this DC component the best fit provides only a $\chi^2$-value of 737 with 370 d.o.f. Values of the DC level are given in Table~\ref{tab:4partpar} in the row labeled ``DC level''.
The DC level might originate from a low energy response of the large \feka line below the escape peak, which is not fully taken into account in the RMF. A similar effect is known to be present in XIS data from \suzaku \citep{matsumoto06a}.

\subsection{Time resolved spectra}
\label{susec:timres_spec}
We have seen in the variability of the light curves, pulsed fractions, and hardness ratios that the overall spectrum will only provide a very inadequate description of the physics of the neutron star, as many different accretion regimes are mixed together. Sacrificing S/N we therefore decided to split the spectrum according to the four regimes I--IV defined in Sect.~\ref{sec:timing} and indicated in Fig.~\ref{fig:lc}. These states show the strongest differences in the light curve and hardness ratio and are therefore expected to show different spectra.

All spectra were rebinned to a S/N of 12 up to 3.5\,keV, of 36 up to 6\,keV, of 60 up to 8.5\,keV and of 48 above. Additionally, at least 4 channels were combined to account for the real energy resolution of the instrument. This binning resulted in different amounts of degrees of freedom for each spectrum. To each of these spectra we fitted the same model as to the overall spectrum and obtained very good \redchi values $\sim$1 for all four spectra. 

The best fit parameters are listed in Table~\ref{tab:4partpar}. It can be seen that the absorption is highest during part II in which the countrate drops very low and the pulsations vanish. During that time the absorption is increased by $\sim$20\% compared to part I. The underlying power law continuum softens at the same time, resulting in comparable hardness ratios. The unabsorbed flux is also highest during part II  (see row $\mathcal{F}_{\text{2--10\,keV}}^{\text{unabs}}$  in Table~\ref{tab:4partpar}), so that the observed drop in count rate can be attributed solely to the very high absorption column.

\begin{table*}
\caption{Fit parameters for the best fit model for the overall spectrum and the spectra of the four states I--IV, respectively. For details on the model see text. Uncertainties are given at the 90\% limit. The two lowest values are obtained when using a gainshift kernel, as discussed in the text.}
\label{tab:4partpar}
\begin{tabular}{r|lllll}
 Parameter & all &  I & II & III & IV \\\hline
Cov. Frac. & $0.935^{+0.005}_{-0.001}$ & $0.943^{+0.011}_{-0.012}$ & $0.9796^{+0.0019}_{-0.0018}$ & $0.89^{+0.02}_{-0.03}$ & $0.968^{+0.010}_{-0.019}$ \\
$N_{\mathrm{H},1}$ [$10^{22}$\,atoms\,cm$^{-2}$] & $141^{+3}_{-2}$ & $147\pm6$ & $175^{+5}_{-4}$ & $114\pm5$ & $130^{+1.2}_{-8}$ \\
$N_{\mathrm{H},2}$ [$10^{22}$\,atoms\,cm$^{-2}$] & $53.7\pm1.6$ & $59\pm4$ & $47\pm2$ & $53^{+3}_{-4}$ & $42^{+9}_{-6}$ \\
Norm [keV$^{-1}$\,s$^{-1}$\,cm$^{-2}$] & $1.15^{+0.00}_{-0.06}$ & $1.19^{+0.18}_{-0.15}$ & $2.2^{+0.3}_{-0.1}$ & $1.02^{+0.09}_{-0.10}$ & $1.1^{+0.4}_{-0.2}$ \\
Photo Index $\Gamma$ & $0.90\pm0.02$ & $0.93\pm0.04$ & $1.10\pm0.04$ & $0.78\pm0.03$ & $1.05^{+0.08}_{-0.03}$ \\
Fe K$\alpha$\,A [ph\,s$^{-1}$\,cm$^{-2}]$ & $\left(1.04^{+0.02}_{-0.03}\right)\times10^{-2}$ & $\left(0.82^{+0.04}_{-0.05}\right)\times10^{-2}$ & $\left(1.10^{+0.03}_{-0.04}\right)\times10^{-2}$ & $\left(1.27^{+0.06}_{-0.07}\right)\times10^{-2}$ & $\left(0.78^{+0.09}_{-0.13}\right)\times10^{-2}$ \\
Fe K$\alpha$\,E [keV] & $6.4227^{+0.0016}_{-0.0001}$ & $6.413\pm0.003$ & $6.407\pm0.002$ & $6.453\pm0.003$ & $6.417^{+0.008}_{-0.006}$ \\
Fe K$\alpha$\,$\sigma$ [keV] & $0.0426^{+0.0016}_{-0.0018}$ & $0.040\pm0.004$ & $0.040\pm0.003$ & $0.032^{+0.003}_{-0.004}$ & $0.029^{+0.008}_{-0.013}$ \\
CS E$_1$ [keV] & $6.205\pm0.009$ & $6.199^{+0.017}_{-0.019}$ & $6.181^{+0.013}_{-0.015}$ & $6.249^{+0.015}_{-0.017}$ & $6.23^{+0.03}_{-0.04}$ \\
CS A [ph\,s$^{-1}$\,cm$^{-2}]$& 
$\left(2.0\pm0.3\right)\times 10^{-3}$ &   $\left(2.4^{+0.4}_{-0.5}\right)\times 10^{-3}$ &   $\left(2.2^{+0.4}_{-0.3}\right)\times 10^{-3}$ &   $\left(3.3^{+0.7}_{-0.6}\right)\times 10^{-3}$ &   $\left(1.3^{+0.7}_{-0.6}\right)\times 10^{-3}$ 
\\
Fe K$\beta$\,A [ph\,s$^{-1}$\,cm$^{-2}]$ & $\left(2.32\pm0.05\right)\times10^{-3}$ & $\left(1.76\pm0.08\right)\times10^{-3}$ & $\left(2.04\pm0.07\right)\times10^{-3}$ & $\left(3.68^{+0.11}_{-0.10}\right)\times10^{-3}$ & $\left(1.53^{+0.15}_{-0.14}\right)\times10^{-3}$ \\
Fe K$\beta$\,E [keV] & $7.095\pm0.002$ & $7.087\pm0.005$ & $7.070^{+0.003}_{-0.004}$ & $7.12669^{+0.00009}_{-0.00016}$ & $7.063\pm0.009$ \\
Fe K$\beta$\,$\sigma$ [keV] & $0.0675^{+0.0011}_{-0.0002}$ & $0.069\pm0.006$ & $0.063\pm0.005$ & $0.052\pm0.005$ & $0.058^{+0.012}_{-0.013}$ \\
Ni K$\alpha$\,A [ph\,s$^{-1}$\,cm$^{-2}]$ & $\left(0.57^{+0.03}_{-0.04}\right)\times10^{-3}$ & $\left(4.4^{+0.5}_{-0.4}\right)\times10^{-4}$ & $\left(0.66^{+0.05}_{-0.04}\right)\times10^{-3}$ & $\left(0.69\pm0.07\right)\times10^{-3}$ & $\left(0.42^{+0.15}_{-0.12}\right)\times10^{-3}$ \\
Ni K$\alpha$\,E [keV] & $7.491\pm0.005$ & $7.484^{+0.005}_{-0.011}$ & $7.473^{+0.004}_{-0.000}$ & $7.515^{+0.015}_{-0.001}$ & $7.50\pm0.03$ \\
Ni K$\alpha$\,$\sigma$ [keV] & $0.036^{+0.010}_{-0.012}$ & $\le0.04$ & $0.053^{+0.011}_{-0.012}$ & $\le0.03$ & $0.06^{+0.05}_{-0.06}$ \\
Ca K$\alpha$\,A [ph\,s$^{-1}$\,cm$^{-2}]$ & $\left(0.73\pm0.06\right)\times10^{-4}$ & $\left(0.47\pm0.09\right)\times10^{-4}$ & $\left(0.82\pm0.09\right)\times10^{-4}$ & $\left(1.03\pm0.18\right)\times10^{-4}$ & $\left(0.8\pm0.2\right)\times10^{-4}$ \\
Ca K$\alpha$\,E [keV] & $3.71^{+0.02}_{-0.01}$ & $3.71^{+0.03}_{-0.00}$ & $3.716^{+0.004}_{-0.011}$ & $3.735^{+0.015}_{-0.009}$ & $3.69^{+0.02}_{-0.03}$ \\
Ar K$\alpha$\,A [ph\,s$^{-1}$\,cm$^{-2}]$ & $\left(3.2\pm0.3\right)\times10^{-5}$ & $\left(1.9\pm0.5\right)\times10^{-5}$ & $\left(3.1\pm0.5\right)\times10^{-5}$ & $\left(0.47\pm0.09\right)\times10^{-4}$ & $\left(0.48^{+0.16}_{-0.14}\right)\times10^{-4}$ \\
Ar K$\alpha$\,E [keV] & $2.973^{+0.012}_{-0.002}$ & $2.969^{+0.016}_{-0.014}$ & $2.955^{+0.015}_{-0.000}$ & $3.000\pm0.015$ & $2.96\pm0.02$ \\
S K$\alpha$\,A [ph\,s$^{-1}$\,cm$^{-2}]$ & $\left(3.3\pm0.4\right)\times10^{-5}$ & $\left(2.1\pm0.5\right)\times10^{-5}$ & $\left(3.0\pm0.5\right)\times10^{-5}$ & $\left(0.61\pm0.10\right)\times10^{-4}$ & $\left(2.9^{+1.7}_{-1.8}\right)\times10^{-5}$ \\
S K$\alpha$\,E [keV] & $2.336^{+0.008}_{-0.004}$ & $2.35\pm0.02$ & $2.315\pm0.011$ & $2.348\pm0.012$ & $2.36\pm0.06$ \\
Abs. Feat.\,A [ph\,s$^{-1}$\,cm$^{-2}]$ & $\left(-0.52^{+0.09}_{-0.10}\right)\times10^{-3}$ & $\left(-3.3^{+1.1}_{-1.2}\right)\times10^{-4}$ & $\left(-2.3^{+0.6}_{-1.9}\right)\times10^{-4}$ & $\left(-1.5\pm0.5\right)\times10^{-3}$ & $\left(-0.8^{+0.3}_{-0.5}\right)\times10^{-3}$ \\
Abs. Feat.\,E [keV] & $5.61\pm0.03$ & $5.57^{+0.06}_{-0.05}$ & $5.67^{+0.03}_{-0.02}$ & $5.70^{+0.00}_{-0.09}$ & $5.56^{+0.10}_{-0.08}$ \\
Abs. Feat.\,$\sigma$ [keV] & $0.26\pm0.03$ & $0.18^{+0.06}_{-0.05}$ & $0.12^{+0.11}_{-0.05}$ & $0.45^{+0.09}_{-0.12}$ & $0.32^{+0.16}_{-0.12}$ \\
Cr K$\alpha$\,A [ph\,s$^{-1}$\,cm$^{-2}]$ & $\left(1.5\pm0.2\right)\times10^{-4}$ & $\left(1.2\pm0.5\right)\times10^{-4}$ & $\left(0.9^{+0.7}_{-0.3}\right)\times10^{-4}$ & $\left(1.9^{+0.6}_{-0.5}\right)\times10^{-4}$ & $\left(1.9\pm0.7\right)\times10^{-4}$ \\
Cr K$\alpha$\,E [keV] & $5.430^{+0.015}_{-0.000}$ & $5.45^{+0.04}_{-0.02}$ & $5.42\pm0.03$ & $5.44\pm0.03$ & $5.44^{+0.02}_{-0.03}$ \\
Ni K$\beta$\,A [ph\,s$^{-1}$\,cm$^{-2}]$ & $\left(1.2\pm0.3\right)\times10^{-4}$ & $\left(0.8\pm0.5\right)\times10^{-4}$ & $\left(1.2\pm0.4\right)\times10^{-4}$ & $\left(1.9\pm0.8\right)\times10^{-4}$ & $\left(0.8^{+0.9}_{-0.8}\right)\times10^{-4}$ \\
Ni K$\beta$\,E [keV] & $8.26^{+0.04}_{-0.02}$ & $8.28^{+0.06}_{-0.08}$ & $8.22^{+0.03}_{-0.04}$ & $8.27^{+0.05}_{-0.04}$ & $8.2^{+0.3}_{-0.2}$ \\
DC Level [keV$^{-1}$\,s$^{-1}$\,cm$^{-2}$] & $\left(0.48\pm0.04\right)\times10^{-4}$ & $\left(3.9\pm0.6\right)\times10^{-5}$ & $\left(3.4\pm0.6\right)\times10^{-5}$ & $\left(0.74\pm0.11\right)\times10^{-4}$ & $\left(0.4^{+0.3}_{-0.2}\right)\times10^{-4}$ \\
Ratio Fe & $0.223^{+0.007}_{-0.006}$ & $0.214^{+0.014}_{-0.013}$ & $0.186^{+0.008}_{-0.007}$ & $0.283^{+0.020}_{-0.016}$ & $0.20^{+0.04}_{-0.02}$ \\
Ratio Ni & $0.20\pm0.05$ & $0.18\pm0.11$ & $0.18\pm0.06$ & $0.23\pm0.14$ & $0.2\pm0.2$ \\
Fe Ions & -- & XII--XIII & VI--XI & $\geq$XIV & IX--XI \\
$\chi^2/\text{d.o.f}\,(\chi^2_\text{red})$ & 448.02/369 (1.214) &  113.69/122 (0.932) & 136.13/120 (1.134) & 129.76/131 (0.991) & 91.72/90 (1.019)\\
$\mathcal{F}_{\text{2--10\,keV}}$ [erg\,sec$^{-1}$\,cm$^{-2}$] & $ 1.043\times10^{-9}$ & $8.863\times10^{-10}$ & $7.127\times10^{-10}$ & $1.852\times10^{-9}$ & $8.048\times10^{-10} $ \\
$\mathcal{F}_{\text{2--10\,keV}}^{\text{unabs}}$ [erg\,sec$^{-1}$\,cm$^{-2}$] & $ 1.617\times10^{-8}$ & $1.624\times10^{-8}$ & $2.151\times10^{-8}$ & $1.797\times10^{-8}$ & $1.350\times10^{-8} $ \\\hline
Gainshift & $1.00396\pm0.00017$ & $1.0023^{+0.0003}_{-0.0004}$ & $1.0010^{+0.0003}_{-0.0002}$ & $1.0082\pm0.0003$ & $1.0019^{+0.0006}_{-0.0005}$ \\
$\chi^2/\text{d.o.f}\,(\chi^2_\text{red})$ & 468.08/375 (1.248) &  129.48/128 (1.012) & 172.64/126 (1.370) & 133.93/137 (0.978) & 101.12/96 (1.053)\\\hline
\end{tabular}
\end{table*}

The width, $\sigma$, of the \feka was in all parts around 40\,eV, larger than the energy resolution of EPIC-pn. If this width is due to Doppler shifts from turbulence in the fluorescent medium, speeds in excess of 1000\,km\,s$^{-1}$ would be necessary. In \gx, observational evidence and models suggest that a wind speed of only 300\,km\,s$^{-1}$ is reached at the position of the neutron star \citep{kaper06a, leahy08a}, which can account for significantly lesser broadening of only $\sim$10\,eV. During accretion the wind will become turbulent, however, and thereby possibly locally increasing the wind speed to values high enough to explain the broadening \citep{mauche08a}. Turbulent flows in the accretion stream are, however, theoretically not yet well understood. A more likely source of apparent line broadening comes from the fact that the \feka is not a single line, but a superposition of different K$\alpha$ lines of differently strongly ionized iron \citep{kallman04a}. This effect can account for a width of 20--30\,eV of the observed line, as the single lines cannot be resolved with the spectral resolution of EPIC-pn. Together with broadening in the wind, the observed width of \feka seems realistic. The \fekb line is even broader with values around 65\,eV. As the \fekb complex is not as broad as the \feka complex \citep{kallman04a}, this broadening might be mainly due to confusion with the strong Fe K-edge at 7.1\,keV, very close to the line. EPIC-pn's resolution does not allow to completely separate these two features, but attempts to fit the spectra with a narrow \fekb line lead to unacceptable \redchi values.

The centroid energy of the \feka line is changing with time and can be as high as 6.45\,keV in part III, where the brightest flare occurred. This energy corresponds to \ion{Fe}{xix} or higher ionized iron, which does not emit a \fekb line, as all M-shell electrons are already stripped from the ion. Still the spectrum of part III shows clear evidence for a \fekb line at  7.127\,keV, corresponding roughly to \ion{Fe}{xiii} to \textsc{xiv} \citep{kallman04a}. 
For the \nika and \nikb lines a similar discrepancy emerges regarding the best fit values: the best fit \nikb centroid energy is very close to the one of neutral nickel \citep{palmeri08a}, while the \nika centroid energy requires at least Ni\,X or higher. The uncertainties of the \nikb energy, however, allow also for ionization states of this order. 

Regarding the flux ratios between the \feka and \fekb  we obtain strongly varying values between
\fekb/\feka = 0.186 -- 0.283. The \nikb/\nika ratio seems to be less variable with values between 0.18 -- 0.23, and the individual uncertainties are much larger so that they are all consistent with each other (see Table~\ref{tab:4partpar}, rows ``Ratio Fe`` and ``Ratio Ni``).
These values are clearly above the theoretical ones for neutral elements from \citet{kaastra93a}, which give \fekb/\feka = 0.125 and \nikb/\nikb = 0.105. Measurements in solid state metals give higher numbers of \fekb/\feka = 0.132 and \nikb/\nika = 0.133 \citep{han09a}. The \nikb/\nika ratio is marginally consistent with all these values within its uncertainties, but the measured iron ratio is clearly higher. Other measurements and calculations for neutral iron give figures of the same order with $\fekb/\feka \approx 0.13$ \citep[see, e.g.,][and references therein]{palmeri03a}. 

It is conceivable that the flourescence lines originate from the dense absorbing material. If this is the case, self-absorption is not negligible \citep{inoue85a}. For an column density of $\nh \approx 140\times10^{22}\,\percmsq$ the optical depth at the energy of the \feka line is only $\approx 60\%$ of the depth at the energy of the \fekb line. If the whole absorption column would act on the lines, the intrinsic ratios are drastically reduced to values between \fekb/\feka = 0.114 -- 0.173, much closer to the theoretical values.  Such a strong effect is, however, unlikely, as the fluorescence lines at softer energies from elements like S and Ar would vanish completely. The line ratio also depends strongly on the viewing angle and the geometry of the absorber \citep[and references therein]{molendi03a}. Simulations have shown that for certain viewing angles around 45$^\circ$ of a dense slab absorber in transmission the measured ratios are reduced, while softer fluorescence lines, especially the Cr~K$\alpha$ line, are still clearly detected (Barrag\'an et al., 2011, in prep.). A detailed investigation will be presented in a forthcoming publication.
For further discussion of the absorption of the fluorescence lines see also Sect.~\ref{susec:flabund}.

The line energies as well as the flux ratio is expected to change with the ionization state of iron \citep{palmeri03a}. Allowing for a mixture of different ionizations during each part and using the energies and yields from \citet{mendoza04a}, we found that ions between \ion{Fe}{vi} to \ion{Fe}{xiv} can explain the observed energies and ratios (see Table \ref{tab:4partpar}, row ''Fe Ions``). Here we neglected self-absorption, as we can not confine its effect on the ratios. Additionally the \fekb line is superimposed on the iron K-edge and the Gaussian model component can be artificially increased due to a correlation with the edge. The real ionization states are therefore likely to be lower, but changes are still clearly present. We find an indication that the ionization state is correlated with the luminosity of the source but that overall the ionization state stays around intermediate values. 
It is furthermore possible
that the energy calibration of \xmm is slightly off and that a small shift in energy depending on the count-rate is present, leading to an apparent shift of the observed line energies.
The most likely effect causing such a behavior is either a change in the charge transfer inefficiency (CTI) of the CCD or a gainshift due to X-ray loading.  The CTI is decreasing with higher photon flux when impurities in the CCD are filled up with electrons \citep{strueder97a}, while X-ray loading causes more electrons than expected. If these changes are not taken into account correctly, they lead to a shift of measured energies depending on the source flux. The expected uncertainties of the calibration of these effects in EPIC-pn are below $\pm0.5\%$ (Guainazzi, 2011, priv. comm.).

To investigate the possibility of such a shift we froze the line energies of all measured elements to the values of their respective neutral line energy, with values taken from \citet{palmeri08b} for S, Ar, and Ca, \citet{kaastra93a} for Cr, \citet{kallman04a} for Fe, and \citet{palmeri08a} for Ni. We then applied a gainshift kernel, shifting the model counts spectrum energy grid according to $E_\text{new}(c) = E_\text{old}(c) / \text{gainshift}$. The obtained best fits were only marginally worse than with free energies of the lines and the shifts were on the order of 1.001--1.008 (0.1--0.8\%), see bottom lines of Table~\ref{tab:4partpar}. These values are only marginally higher than expected from calibration uncertainties but also include unknown source intrinsic energy shifts. The continuum parameters of the spectral model did not change significantly due to the gainshfit, but the fluxes of the fluorescence lines were slightly altered to values marginally consistent with the ones shown in Table~\ref{tab:4partpar}. Especially the \fekb line flux was decreased, possibly due to a better seperation from the iron K-edge. This effect influenced also the \fekb/\feka ratio by decreasing it slightly to values $\leq$$0.2$ at all times, which is closer to the values expected for neutral iron. No significant change with luminosity could be observed.

Using the gainshift kernel we can explain the measured line energy shifts stemming
 purely from detector effects while the physical condition of the plasma in \gx does not change significantly during our observation.

\subsection{Pulse to pulse spectroscopy}
\label{susec:pulse_spec}

\begin{figure}
 \centering
 \includegraphics[width=0.95\columnwidth]{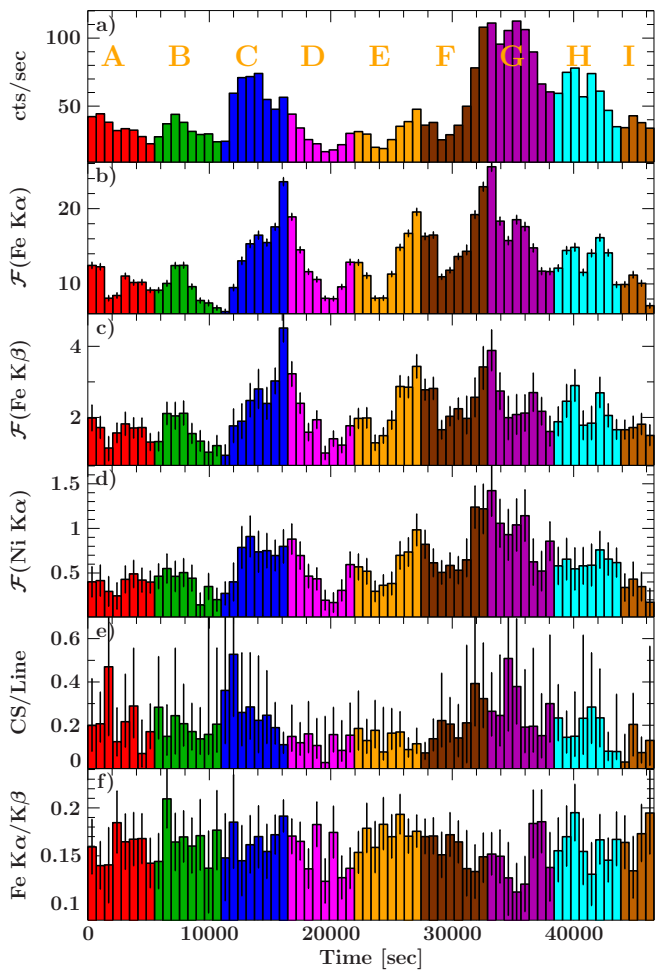}
 \caption{Evolution of the spectral parameters of individual pulses. \textit{a)} Light curve between 0.5--10\,keV , \textit{b)} \feka flux (including the Compton shoulder), \textit{c)} \fekb flux, \textit{d)} \nika flux, \textit{e)} ratio between the Compton shoulder and the Gaussian shaped \feka line, and \textit{f)} ratio between the \feka and the \fekb flux. The fluxes of the lines are given in photons\,s$^{-1}$\,cm$^{-2}\times10^{-3}$.}
 \label{fig:lc_lines}
\end{figure}

\begin{figure}
 \centering
 \includegraphics[width=0.95\columnwidth]{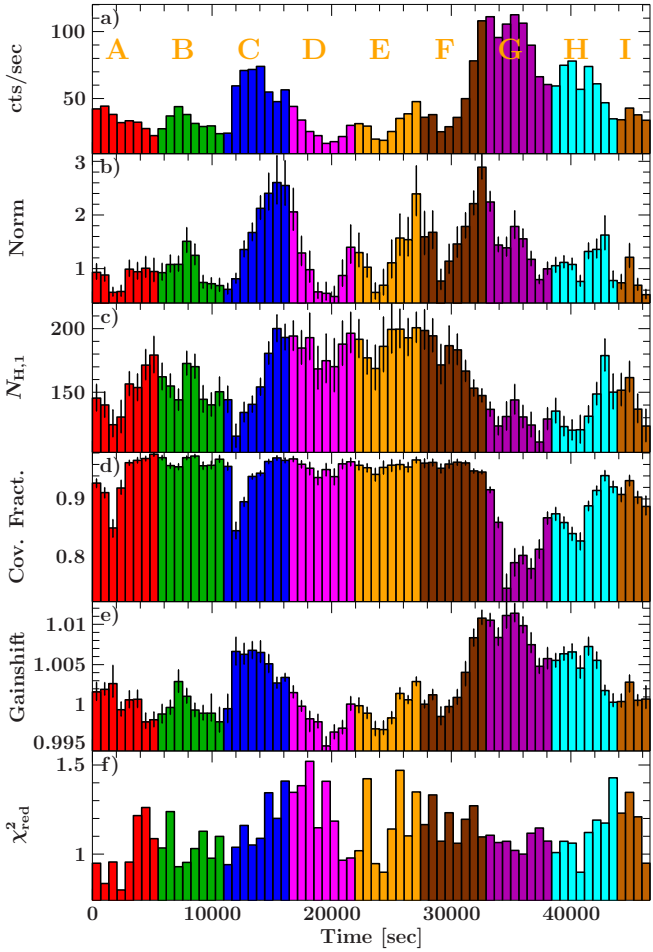}
 \caption{Time series of the spectral parameters of the spectra of individual pulses. \textit{a)} Light curve between 0.5--10\,keV,
\textit{b)} continuum norm in photons\,keV$^{-1}$\,s$^{-1}$\,cm$^{-2}$, \textit{c)} main absorption column in $10^{22}$\,\atmcmsq, \textit{d)} covering fraction, \textit{e)} gainshift slope, and \textit{f)} \redchi of the best fit.}
 \label{fig:lc_cont}
\end{figure}

In an approach to understand the time variability of the spectral parameters, especially the iron line complex, in more detail, we extracted spectra for individual pulses, leading to 69 spectra with 685.0\,sec exposure time each. The resulting model parameters are presented in Figs.~\ref{fig:lc_lines} and \ref{fig:lc_cont}. The data are grouped by color into 9 parts labeled A--I to facilitate orientation. We applied a reduced version of the time averaged spectral model and included only the \feka line with its Compton shoulder, the \fekb, and the \nika line. All other fluorescent lines could not be identified in these short exposure spectra. Additionally we had to fix the photon index $\Gamma = 0.90$ and the secondary absorption column $\nhtwo =5.37\times10^{23}\,\atmcmsq$ because neither could be constrained well with the data. When leaving the centroid energy of the emission lines free, we found a strong correlation between these energies and the average count rate, as already indicated in the previous Section. This correlation is clearly stronger than expected from ionization \citep{kallman04a} and is therefore likely a detector effect.
 As described in the previous Section, these calibration uncertainties can be modeled with a gainshift kernel. We also checked the spectra extracted using \texttt{epfast}, but they showed the same systematic effects.

With fixed energies for the fluorescent lines and for the iron K-edge the use of a gainshift kernel leads to a very good  description of the individual spectra. As expected the gainshift is directly correlated to the incident countrate, see Fig.~\ref{fig:lc_cont}\textit{e}.

The flux of the \feka line and its Compton shoulder is mostly, but  not always, correlated with the count rate in the 0.5--10\,keV energy band, as seen in Fig.~\ref{fig:lc_lines}. Especially in the second, main flare (part G) the iron line flux starts to rise with the light curve but breaks down around the maximum of the flare and stays low for the remainder of the observation. A better correlation is observed with the normalization of the power law continuum (Fig.~\ref{fig:lc_cont}). The \fekb and the \nika lines show a very similar behavior and are strongly correlated with the \feka line flux, with a Pearson's correlation coefficient of 0.92 and 0.89, respectively. As discussed in the previous Section, the \feka and \fekb lines required a finite width $\sigma$, with an average value of 40\,eV and 71\,eV, respectively. The widths did not show a significant correlation with time or flux. The width  of the  \nika was frozen to $10^{-4}$\,keV. 

The absorption column is also strongly changing over the course of the observation, varying by more than 50\% (see Fig.~\ref{fig:lc_cont}\textit{c}). It is increasing during the first flare in part C and stays high until the onset of the second, main flare (parts F and G), where a strong drop is registered. After the flare it is increasing again gradually. The covering fraction is very high throughout the observation, with an average of $\sim$92\%, reducing the secondary absorption column to almost negligible values. It is only during the bright flare in part G where the covering fraction suddenly drops as low as 74\%, around the time when the \feka line flux dips.

\begin{figure}
 \centering
 \includegraphics[width=0.95\columnwidth]{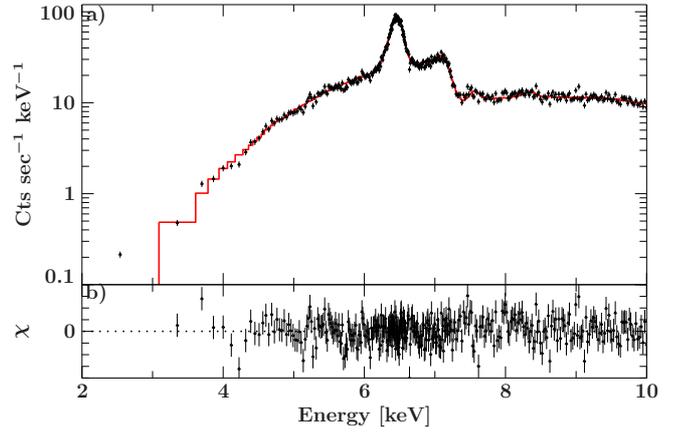}
 \caption{Typical spectrum with 685.0\,sec exposure time, i.e., one pulse. The upper panel shows the data and the best fit spectrum, the lower panel shows the respective residuals. For details about the spectral model see text.}
 \label{fig:spec_p48}
\end{figure}

A typical spectrum with one pulse exposure time is shown in Fig.~\ref{fig:spec_p48}. The \feka and \fekb lines as well as the iron K-edge, are clearly seen. The \nika line is also visible on close inspection, but below 3\,keV no significant excess over the background can be measured anymore due to the huge absorption column. The Ca, Ar, S, and Cr K$\alpha$ lines and the \nikb line are not significantly detected.

\begin{figure}
 \centering
 \includegraphics[width=0.95\columnwidth]{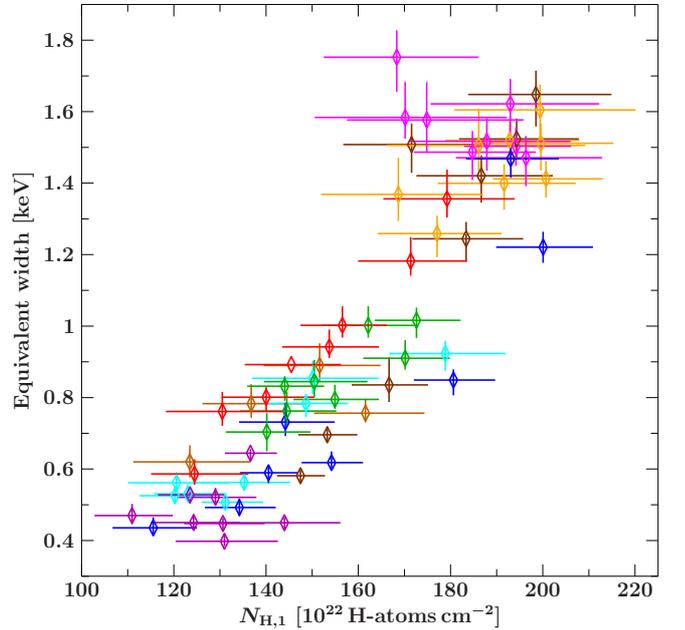}
 \caption{Equivalent width of the \feka line as function of the main absorption column \nhone. Colors are the same as in Fig.~\ref{fig:lc_cont} and are indicating the time.}
 \label{fig:nh2eqw}
\end{figure}

Assuming that the absorbing matter is the same matter responsible for the fluorescent iron emission and that it is distributed spherically symmetric around the X-ray source, the equivalent width of the \feka line is proportional to the overall absorbing column \citep{inoue85a}. Furthermore, self-shielding should be negligible and the matter is assumed to be cold and neutral. We can see this behavior during most of the \xmm-observation, even at very large \nh values, see Fig.~\ref{fig:nh2eqw}. The average Pearson's correlation coefficient over all parts except part D is $0.91$. During part D, corresponding to the off-state seen in the light curve (Figs.~\ref{fig:lc} and \ref{fig:vanishedpulse}), this correlation is not visible. Instead indication for an anti-correlation with a correlation coefficient of $-0.67$ exists.
During that state the equivalent width of the \feka line reaches its highest values, whereas the \nh stays in a regime around 1.8$\times 10^{24}$\,\atmcmsq, lower than the maximal measured value. The overall \nh value is quite high, as expected from the analysis of region II from Sect.~\ref{susec:timres_spec} (corresponding to parts D, E, and F) which showed the highest overall absorption value. But the off-state is happening during part D only, a part which does not stand out for its increased absorption column when compared to parts E and F. This invariance of the \nh value clearly shows that the off-state is not due to a increased absorption, which was already indicated by the softening during that state (see Fig.~\ref{fig:lc}).

\section{Discussion \& Conclusion}
\label{sec:conlusion}

We have presented a detailed timing and spectral analysis of the HMXB \gx during its pre-periastron flare using the EPIC-pn camera of \xmm. The 2--10\,keV spectrum can be described with a partially covered, strongly absorbed power law model enhanced by several Gaussian emission lines from S, Ar, Ca, Fe, and Ni. We found evidence that the K$\alpha$ line from Cr is also present as well as the K$\beta$ line from Ni. To our knowledge this is the first detection of these lines in a HMXB. In order to account for the different states of activity visible in the light curve we separated the data according to these states, labeling them I--IV. During the sub-flares (parts I and III) regular pulsations with  $P = 685.0$\,sec  were clearly detected and a double peaked pulse profile is visible. During the off-state II these pulsations vanished almost completely and the countrate dropped to very low values of $\sim$10\% of the peak count rate. \xmm still detected the source clearly.  
By separating the data into individual spectra for every rotation of the neutron star, we were able to perform a detailed spectral analysis of the off-state. We found that although the absorbing column is very large, it is not drastically increased during that state, compared to values measured shortly before and after it. Instead a softening of the spectra is observed in which the \feka line is even more important than at other times.

We have shown that the absorption column is larger than $10^{24}\,\atmcmsq$ at all times during the observation. \gx is not the only source showing absorption columns of this magnitude. \inte, for example, discovered more than 10 highly absorbed HMXB, which might form a new class of objects \citep{kuulkers05a,walter06a}. Among them are sources like IGR~J16318$-$4848 \citep{matt03a, barragan09a} or IGR~J17252$-$3616 \citep{manousakis11a} which show very prominent \feka and \fekb lines, as well as \nika lines. The flux of these sources is, however, usually lower than that of \gx, decreasing the S/N drastically below 4\,keV and thereby not allowing for studies of softer fluorescence lines. Results from the iron line energies and \feka/K$\beta$ ratios also indicate only mildly ionized iron. \gx could therefore be a brighter example of these sources and help to understand this class of HMXB.

\subsection{Fluorescence region size}
\label{susec:fluor_size}
To estimate the distance between the X-ray producing region and the fluorescent medium, we extracted light curves with 1\,sec resolution for the iron line energy band between 6.3--6.5\,keV and the harder continuum band between 7.3--8.5\,keV. The countrate between 6.3--6.5\,keV is completely dominated by the \feka line and any variability of it must be due to the reaction of the fluorescent line to changes in the continuum flux. We compared both light curves using Fourier transformation as well as cross-correlation techniques and could not find a significant time delay between both curves in the range of 2--5000\,sec. 
This is in agreement with the size of the fluorescence region being smaller than $10^{10}$\,cm, i.e., 0.3\,lt-sec, as given by \citet{endo02a}. The count rate in the narrow energy bands was not high enough to achieve a time resolution sufficient to see a delay of the fluorescence lines on that order.  For IGR~J17252$-$3616 \citet{manousakis11a} also found that the emission line region is compact and not larger than the accretion radius. However, during the off-state in part D in Fig.~\ref{fig:lc_lines} the iron line is still clearly visible, although the luminosity of the continuum has significantly decreased. This leads to the largest equivalent widths for the \feka line measured in the whole observation (Fig.~\ref{fig:nh2eqw}) and seems to indicate that the fluorescence line is partly reacting rather slowly on changes in the continuum. A possible explanation is that the weaker part of the partial coverer is located further out,
 and thereby reacting much slower to any changes in the X-ray continuum flux. 
The X-ray flux was low for $\sim$4500\,sec and we can assume this scale as an upper limit of the light travel time to the absorber. On a direct path, this would place the absorbing medium almost 10 times the average orbital diameter of the system away. Through scattering in the dense medium, with an optical depth $\tau = \nhone\times\sigma_\mathrm{T} \approx 1$, the effective light-path is increased, so that a size of 2000\,lt-sec or less is plausible. 
A similar geometry was already proposed by \citet{mukherjee04a}. Through scattering in the inner absorber the outer absorber receives a much less pulsating X-ray flux and transforms it into fluorescence lines, which are consequently also pulsating much less. This effect can explain the reduced pulsed fraction in the \feka and \fekb line band (Fig.~\ref{fig:pulfracenerg}). A secondary absorber further out can also explain the dip in \nh and covering fraction during the large flare in part G (Fig.~\ref{fig:lc_cont}). Here the inner absorber is almost completely ionized and transparent to soft X-rays. The visible absorption is then primarily due to the still cold outer absorber. The present data do not allow one to distinguish between the lines formed in the different absorbers, and as they are both variable it is hard to disentangle the effects. The indications we obtain from the data do, however, support this absorption geometry. If both columns of the partial covering model would be located close to the neutron star, where the strong column describes the clumpy stellar wind and the other the smooth part of the wind, neither the large equivalent width nor the strong dip in covering fraction during the large flare is easily explainable. The measured widths of the \feka and \fekb lines as well as their ratio (Sec.~\ref{susec:timres_spec}) can also be understood in this geometry as a superposition of cold and neutral iron in the outer absorber and higher ionized iron in the inner one. As stated before, the spectral resolution of EPIC-pn does not allow to separate different iron lines, but we would expect to see an only slowly variable \feka line of neutral iron and many variable lines from the different ionized species from the inner absorber.

\subsection{Fluorescence and abundances}
\label{susec:flabund}
The 2--10\,keV spectrum of \gx shows evidence for \feka, \fekb, \nika, \nikb, S\,K$\alpha$, Ca\,K$\alpha$, Ar\,K$\alpha$, and Cr\,K$\alpha$ fluorescent emission lines. 
 To check consistency with older data, especially regarding the Cr\,K$\alpha$ line,  we applied our model to the \chandra HETG data presented by \citep{watanabe03a} and taken during the pre-periastron flare in January 2001. The model describes the data very well in its overall shape. Neither a Cr\,K$\alpha$ nor a \nikb line is significantly detected, but lines with the same flux and width as in the \xmm spectrum are also not ruled out by the data and could vanish due to the worse S/N of the \chandra data.

To describe the \xmm spectrum we used a model in which the fluorescent lines were only absorbed by the galactic absorption column of $\nh^{\text{gal}}=1.7\times10^{22}\,\atmcmsq$ to accommodate for the observed fluxes of the K$\alpha$ lines of S and Ar, see Sect.~\ref{susec:avg_spectra}. The flux of the lines indicates, however, that they originate in a dense absorber and that thus self-absorption must be taken into account. Additionally timing analysis has shown that the main producing region of the iron line must be very close to the X-ray source and it is likely that this is true for the other lines as well. It is thus conceivable that the photons emitted from the fluorescent ions are also absorbed by material relatively close to the neutron star with a column density in excess of $\nh^{\text{gal}}$, e.g., by the stellar wind of the companion. 

In order to estimate how strongly these lines are absorbed we performed the following simple sanity check, assuming solar abundances for all elements \citep{wilms00a}: for each element, where a fluorescent line was observed, we calculated how many photons were absorbed by this element alone above its respective K-shell, applying the cross-sections by \citet{verner96a}. Using the fluorescence yields for near neutral ions from \citet{kaastra93a} we calculated how many of these photons were reemitted in the respective K$\alpha$ and K$\beta$ lines. As we do not know the geometry of the fluorescent medium and hence can not tell how many of these reemitted photons reach us, we calculated the fluxes of all lines relative to \feka. We compared these relative fluxes to the fluxes as measured by the Gaussian lines in our model. The relative fluxes of the Gaussian lines are strongly correlated with the absorption column applied, especially for the soft S and Ar lines. By changing the absorption column we can find the best agreement between measurement and calculation. We find that this is achieved with an overall absorption column around $20\times10^{22}$\,\atmcmsq, i.e., a factor of $\sim$6 less than the continuum absorption and consistent with the above assumptions. Systematic uncertainties outweigh any statistical uncertainties in this analysis and are estimated to be on the order of $\pm 5\times10^{22}\,\atmcmsq$. A drastic improvement on this estimate can be achieved using Monte Carlo simulations of different absorber geometries, but this investigation is beyond the scope of this paper (see Barrag\'an et al., 2011, in prep.).

\subsection{The chromium line}
We have reported the detection of an emission line at 5.43\,keV, compatible with the K$\alpha$ line energy of neutral chromium. This line is weak and only visible because of the large absorption column of \gx  making it bright enough to be visible over the continuum. The highly absorbed IGR sources described above are typically not bright enough to provide enough S/N to allow searches for a Cr line. In Supernova Remnants (SNR) evidence for fluorescent chromium K lines was found, most notably in W49B \citep{hwang00a, miceli06a} but also in Tycho, Cas A, and Kepler \citep{yang09a}. Cr is a decay product of Fe in supernovae, and thus a high iron abundance should be accompanied by a high chromium abundance. The abundance ratio of Cr to Fe is $\sim$1\% for solar mass distribution. In supernovae the ratio depends on the explosion model and can be 2\% or larger for delayed detonation models \citep[see, e.g.,][]{iwamoto99a}. The SNRs investigated by \citet{yang09a} show different abundance ratios between 0.63\% -- 3.6\%, leading the authors to suspect that they originate from different type of supernovae explosions. 
Using the method described in the previous Section to calculate the relative strengths and therewith the relative abundances of Cr to Fe we obtain a value of 1.4\%, very well within the expected values. Taking only the ratio of the measured photons in the Cr\,K$\alpha$ and \feka line, we obtain higher values of $\sim$6.4\%, but this value neglects the fact that the K-edge for Cr is lower than the one of Fe and that consequently more photons are able to ionize Cr. With our measurement we can not infer possible explosion models or production channels of the \gx system. The measured relative line flux does, however, agree with predicted fluorescence yields and element abundances.

\subsection{Accretion during the off-state}
The off-state is perhaps the single most puzzling feature of the \xmm data. A similar off-state was only recently discovered by \citet{goegues11a} in \xte data. \xte's spectral resolution does not allow for a detailed analysis of the iron line region but provides a larger broad band coverage compared to \xmm. \citet{goegues11a} fitted individual spectra with exposure times of one pulse period each, keeping the iron line fixed and letting the photon index of the power law continuum vary. They found that the photon index is drastically increased during the dip, making the overall spectrum much softer. It thus seems very likely that the accretion was switched off during that time and that only weak Comptonization in the accretion columns was possible, while most of the X-ray flux originated from the black body on the neutron star's surface. This result is in accordance with a theory developed to explain similar off-state in Vela~X-1 \citep[and references therein]{kreykenbohm08a}. According to this theory, accretion ceases during the off-state and the source enters the propeller-regime, i.e., a regime in which the Alvf\'{e}n radius is larger than the co-rotating radius \citep{lamb73a}. This transition is triggered by a strong decrease in the mass accretion rate, possibly due to a strongly structured or clumped wind \citep{oskinova07a, kreykenbohm08a, fuerst10a}. Even though \gx is not  directly accreting from the wind in the pre-periastron flare but rather from an accretion stream \citep{leahy08a}, strong density variations are evident through the flaring behavior of the light-curve (Fig.~\ref{fig:lc}). Other observational evidence for a clumpy wind around Wray~977 was put forward by \citet{mukherjee04a} who studied in detail the time variability of the absorption column on short and long timescales. 

All our results indicate that the  observed off-state was caused by a cessation of accretion and not by an optically-thick clump moving through the line of sight. A clump in the line of sight would also lead to drastically reduced pulsations and flux, as the X-ray flux is scattered in the absorber and thereby smearing out the pulsations \citep{makino85a}. In the spectral analysis, however, we would expect that the spectrum gets harder, as softer X-rays are affected stronger by absorption than harder X-rays. We would  also expect to see a drastic increase of the absorption column. Both are not the case, as seen in Fig.~\ref{fig:lc_cont}. 

\begin{figure}
 \centering
 \includegraphics[width=0.95\columnwidth]{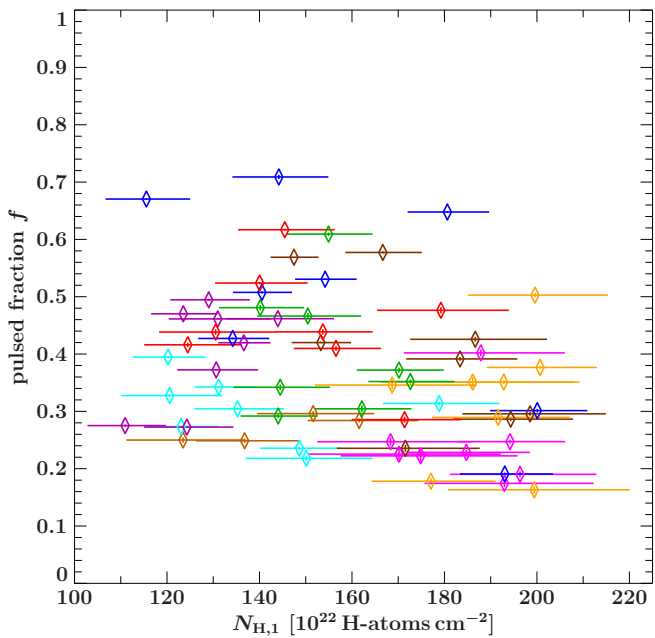}
 \caption{Pulsed fraction as function of \nhone. Error bars on the pulsed fraction are arbitrary. Color code is the same as in Fig.~\ref{fig:lc_cont}.}
 \label{fig:pulfrac2nh}
\end{figure}

We investigated the overall behavior of the pulsed fraction as function of \nh, and found almost no correlation (Fig.~\ref{fig:pulfrac2nh}), contradictory to the results by \citet{makino85a}. Scattering in the absorption medium seems to be influencing the pulsed flux only marginally, so that the complete disappearance of pulses in part D is very unlikely to be caused by larger \nh values. The variation in the pulsed fraction seems to originate from different accretion rates, as expected from a clumpy accretion medium. This influence is also supported by the variability of the individual pulse profiles, as shown in Fig.~\ref{fig:ppex}. We do not see a weaker pulsed flux with essentially the same shape causing the lower pulsed fraction, but rather a clear change in the overall pulse profile caused by a different accretion rate.

 From results at harder X-rays \citep{evangelista10a} a decrease of the pulsed fraction with decreased luminosity would be expected. In a simple toy model this can be explained by a change in altitude of the X-ray producing region in the accretion column \citep{lutovinov09a}. Such a change would also be visible in a change of the energy of the CRSF \citep{suchy11a, doroshenko10a}.  We did not find a correlation between the pulsed flux and the luminosity in the \xmm data. The pulsed fraction during the pre-periastron flare is the lowest over the whole orbit and is also much lower in the soft than in the hard X-rays \citep{evangelista10a}. It is thus possible that a change in the accretion column might not be visible in the pulsed fraction of the 2--10\,keV band.

Assuming that the off-state was caused by a cessation of accretion, possibly because the neutron star entered the propeller regime, the remaining X-ray flux is mainly coming from the still fluorescent absorbing medium around the neutron star, as described in Sec.~\ref{susec:fluor_size}.

As shown in Fig.~\ref{fig:nh2eqw} during the normal states the equivalent width of the \feka is correlated with the \nhone value. This correlation is explained by the assumption that the fluorescent medium is the same as the absorbing medium. As a result, a higher absorption column leads also to a  brighter fluorescence line as more atoms in the absorber are ionized by the X-ray continuum. During the off-state, the X-ray continuum is drastically reduced leading to a reduced \feka line from the main absorber in the line of sight. The off-state indicates a very clumpy medium, so that the absorption in the line of sight is not necessarily an indicator for the overall density of the medium surrounding the X-ray source. The observed break-down of the correlation between equivalent width of the \feka line and the \nhone value can thus be explained by a complex structure of the medium around the neutron star. For example, fluorescence  from dense medium behind the neutron star could arrive simultaneously with the reduced continuum flux at the observe due to different light-travel times.
The remaining continuum flux during the off-state could also be due to Compton reflection from a dense medium around the neutron star, similar to the scenario proposed by \citet{rea05a} to explain a long low-intensity state in the Low Mass X-ray Binary (LMXB) GX~4+1. These authors analysed broad band \sax spectra  and found that they could be described by a cut-off power law plus a reflection component, the latter becoming more important during the low intensity state. In this state also the \feka line reached the hightes equivalent widths, similar to \gx. \citet{rea05a} found that although the pulsations vanished below 7\,keV they could still be seen in the harder X-rays, an effect we can not investigate with the \xmm data but which is not evident in the \xte/PCA data presented by \citet{goegues11a}. If the reflecting surface is not as well defined as in the geometry proposed for GX~1+4, all pulsations can be smeared out over a broad energy range, so that the reflected continuum flux is also unpulsed when escaping the medium around the system. More observations in a broader energy range are necessary to understand the geometry further.

\subsection{Outlook}
Off-state are seen in an increasingly number of sources, like Vela~X-1 and GX~1+4 and can provide interesting insight in the structure of the accreted material as well as the configuration of the accretion and absorption region. For \gx we propose that two absorbers exist, a very dense but clumpy primary one close to the neutron star, and a weaker and smoother secondary one farther away. Fluorescence lines from the secondary absorber are much weaker then the ones from the primary and only visible during the off-state. 
High-resolution spectra of the off-states from \chandra HETG or future missions will help to investigate them in more detail by allowing for better measurements of
the ionization state 
and the geometry of the absorber.
The discovery of a Cr~K$\alpha$ line in the spectrum of \gx leads to interesting new questions about the abundances of the iron family atoms, including Cr and Ni. With higher effective areas in future missions, more sources might be found showing a Cr~K$\alpha$ line and in comparing these findings to \gx more insight into the production of HMXB and the preceding supernova can be achieved. With broad-band spectroscopy, as already possible with the \suzaku observatory, the underlying continuum can be constrained simultaneously in a better way than with \xmm oder \chandra only. Our collaboration is analyzing \suzaku data of \gx taken during orbital phases outside the pre-periastron flare \citep{suchy11a}. These observations provide insight into the behavior of the accretion column through exact measurements of the variability of the CRSF energy. All this information combined will complete the picture of \gx and possibly also of the other highly absorbed sources.

\acknowledgements
We thank the anonymous referee for her/his useful comments. This work was supported by the Bundesministerium f\"ur Wirtschaft und Technologie through DLR grant 50\,OR\,0808, via a DAAD fellowship, and has been partially funded by the European Commission under the 7th Framework Program under contract ITN\,215212 ``Black Hole Universe''. SS acknowledges support by NASA contract NAS5-30720 and NNX08AX83G. IC acknowledges financial support from the French Space Agency CNES through CNRS. FF thanks the colleagues at UCSD and GSFC for their hospitality. We have made use of NASA's Astrophysics Data System. This work is based on observations obtained with \xmm, an ESA science mission
with instruments and contributions directly funded by ESA Member States and NASA.
We made use of results provided by the ASM/\xte teams at MIT and at the \xte SOF and GOF at NASA's GSFC. We used the ISIS software package provided by MIT for this work. We especially like to thank J. C. Houck and J. E. Davis for their restless work to improve ISIS and S-Lang.

\end{document}